\documentclass[preprint,pre,epsfig,address]{revtex4}%
\usepackage[english]{babel}
\usepackage[latin1]{inputenc}
\usepackage[dvips]{graphicx}
\usepackage{amsmath}
\usepackage{amssymb}
\usepackage{epsfig}
\usepackage{array}
\usepackage{amsfonts}
\usepackage{graphicx}%
\begin{document}
\title{ Why spontaneous symmetry breaking disappears in a bridge system with
PDE-friendly boundaries }
\author{Vladislav Popkov}
\altaffiliation{Present address :
Institut f\"ur Theoretische Physik, Universit\"at zu K\"oln, Z\"ulpicher Str. 
77, D-50937 Cologne, Germany}

\author{Gunter M. Sch\"utz}
\affiliation{ Institut f\"ur Festk\"orperforschung,
Forschungszentrum J\"ulich, D-52425 J\"ulich, Germany }

\date{\today }

\begin{abstract}
We consider a driven diffusive system with two types of particles, A and B,
coupled at the ends to reservoirs with fixed particle densities. To define
stochastic dynamics that correspond to boundary reservoirs we introduce
projection measures. The stationary state is shown to be approached
dynamically through an infinite reflection of shocks from the boundaries. We
argue that spontaneous symmetry breaking observed in similar systems is due to
placing effective impurities at the boundaries and therefore does not occur in
our system. Monte-Carlo simulations confirm our results.

\end{abstract}
%\pacs{87.14.Ee, 87.15.Aa, 87.15.Vv}
\maketitle

\section{Introduction}

Systems of driven diffusing particles attract attention because, despite their
relative simplicity, they embrace a whole range of critical phenomena far from
thermal equilibrium \cite{Schu00,Liggett1999,Zia}. One of the
remarkable features is the appearance of phase transitions induced by spatial
boundaries of a system, studied in detail for models with one species of
particles \cite{Krug91,ASEP1,ASEP,Kolo98,Gunter_Slava_Europhys} and for 
some multi-species models
\cite{Mukamel95,Peschel,Erel,Rako04a}. The ability of a nonequilibrium
system to \textquotedblleft feel\textquotedblright\ the boundaries constitutes
a key feature of driven systems: in fact, the boundaries dominate the bulk
giving rise to the phase transitions. Therefore it is clear that boundary
conditions play a crucial role and have to be chosen with care, strictly
adequate to the physical situation which is being modelled. To this end one
has to define specific boundary rates for injection and extraction of
particles at the ends of the system. There are several possible strategies.

One of them is to postulate the most simple boundary rates, and treat the
rates themselves as parameters. A system is studied, then, as a function of
those parameters, so that one tries to keep their number small. This approach
is used, e.g., in \cite{Krug91,Mukamel95}.

Another approach, let us call it \textit{particle boundary reservoir
approach}, treats the boundary problem of a driven diffusive system as if it
was coupled at the ends to reservoirs of particles with fixed particle
densities. In this case, the rates are \textit{dictated} \ by the reservoir
particle densities and by themselves do not, in general, look simple. The
boundary rates are fixed such that the stationary measure of the process with
equal reservoir densities on the left and on the right is essentially
equal to the stationary measure of the infinite system with the same particle
densities, see below for a detailed discussion. Thus reservoir particle
densities are parameters, namely, reservoir densities of each species of
particles at each boundary. Correspondingly, in the one-dimensional case,
where there are two ( left and right) boundaries, the number of parameters is
equal to $2\ast number$ $of$ \ $species$ in the system. This approach is used
in \cite{Gunter_Slava_Europhys,Peschel,TiborAntal}. We argue
that in this case there exists a well-defined hydrodynamic limit, and the
original stochastic problem has correspondingly a well-behaved coarse-grained
description via conservation law equations. The latter constitutes a powerful
tool to study the physics of the system, e.g., phase diagrams and the location
and type of phase transitions.

In some important cases the two approaches are equivalent. For instance, for
the ASEP with open boundaries, the rates of injection $\alpha$ and extraction
$\beta$ respectively are expressed as $\alpha=\rho_{L},\beta=1-\rho_{R}$ in
terms of the particle boundary reservoir approach. In more complicated settings
one may attribute effective boundary densities to given boundary rates.
In this way a very large class of single-species systems with open boundaries can be described
and their phase diagram can be predicted in terms of the boundary densities
\cite{Gunter_Slava_Europhys}. For attractive single-species systems the validity of this 
approach has been proved rigorously in very recent work \cite{Baha04}.
The purpose of this paper is to demonstrate that in two-species models, however, 
the two approaches cannot generally be substituted with one
another. %An example where both are strikingly different, are systems with two
%species of particles. 
We exhibit this difference and discuss its
significance for the appearance of spontaneous symmetry breaking in an ABO model
\cite{Mukamel95}, which is a paradigmatic two-species model \cite{Gunter_reviewJPA}.

To this end, we first introduce the model (Sec. II), discuss its hydrodynamic
properties under Eulerian scaling (Sec. III) and show how to construct
appropriate boundary reservoirs (Sec. IV). The strategy employed here can be applied
to any stochastic particle system. Then we
show (Sec. V)  that the phenomenon of
infinite reflections of shock fronts from the boundaries
\cite{reflections_JPA} exists also in the ABO system which is much studied,
and is known to have many intriguing features. Using this we arrive at the
conclusion (Sec. VI) that the
so-called bridge model \cite{Mukamel95} (which is an ABO model with specific
injection and extraction rates) has spontaneous symmetry breaking due to
the fact that the boundary injection and extraction rates place effective
obstacles at the ends of the chain, rather than corresponding to boundary 
reservoirs with some effective boundary density.
We demonstrate that with these effective obstacles removed, there is no 
symmetry breaking phase transition. Thus,
symmetry breaking can be viewed as an impurity-induced effect. Moreover, we
argue that the original bridge model may have symmetry breaking transition
precisely at the point where a hydrodynamic description 
with effective boundary densities breaks down. We summarize our main conclusion
in the final Sec. VII.

\section{The model}

The ABO model is defined on a chain of length $N$. There are particles of type
$A$ and of type $B$, distributed along the chain. Any site of the chain can be
empty, or contain at most one particle. In the bulk, exchange with the nearest
neighbours happens, e.g., a particle $A$ can exchange with a hole $0$ on its
right $A0\Rightarrow0A$ with rate $\Gamma_{A0}$, etc.. The complete set of
rates is given as follows%
\begin{align}
A0  &  \Rightarrow0A\text{ \ \ \ \ \ \ with rate }\Gamma_{A0}\nonumber\\
0A  &  \Rightarrow A0\text{ \ \ \ \ \ \ with rate }\Gamma_{0A}\nonumber\\
B0  &  \Rightarrow0B\text{ \ \ \ \ \ \ with rate }\Gamma_{B0}\label{rates}\\
0B  &  \Rightarrow B0\text{ \ \ \ \ \ \ with rate }\Gamma_{0B}\nonumber\\
AB  &  \Rightarrow BA\text{ \ \ \ \ \ \ with rate }\Gamma_{AB}\nonumber\\
BA  &  \Rightarrow AB\text{ \ \ \ \ \ \ with rate }\Gamma_{BA}\nonumber
\end{align}
The boundary conditions correspond to reservoirs of particles with fixed
densities $\rho_{L}^{A},\rho_{L}^{B}$ and $\rho_{R}^{A},\rho_{R}^{B}$ at the
left and at the right end of the chain respectively and are defined in the
section IV. For a specific choice of rates \cite{Rittenberg98},
\begin{equation}
\Gamma_{AB}-\Gamma_{BA}=(\Gamma_{A0}-\Gamma_{0A})+(\Gamma_{0B}-\Gamma_{B0}),
\label{product_measure_condition}%
\end{equation}
the stationary state on a ring is given by a product measure, which does not
crucially affect the qualitative properties of the system, but significantly
reduces computational difficulties (since no approximation in the calculation
of the bulk flux, boundary conditions are required). We shall follow this
choice below, making comments on the general situation when necessary. In most
cases, we shall take%
\begin{equation}
\Gamma_{A0}=\Gamma_{0B}=\Gamma_{AB}/2=1,\text{ \ \ \ }\Gamma_{0A}=\Gamma
_{B0}=\Gamma_{BA}=0\text{\ \ } \label{rates112}%
\end{equation}
without losing generality.

\section{Hydrodynamic limit equations}

Here we shall treat for simplicity the case (\ref{rates112}) where there is a
product measure state on a ring. From this property and the hopping rules
(\ref{rates112}) exact stationary fluxes are given by
\begin{align}
j^{A}(\rho^{A},\rho^{B})  &  =\rho^{A}(1-\rho^{A}+\rho^{B})\label{fluxA}\\
j^{B}(\rho^{A},\rho^{B})  &  =-\rho^{B}(1-\rho^{B}+\rho^{A}). \label{fluxB}%
\end{align}
Following \cite{GunterSlava_StatPhys}, the hydrodynamic equations of our
process may be obtained by averaging, factorizing and Taylor expanding the
lattice equations of motion with respect to the lattice constant $a\ll1$ . In
the first order of the expansion, one obtains%
\begin{align}
{\frac{\partial\rho^{\mathrm{A}}}{\partial t^{\prime}}}+{\frac{\partial
j^{\mathrm{A}}(\rho^{\mathrm{A}},\rho^{\mathrm{B}})}{\partial x}}  &
=\varepsilon{\frac{\partial}{\partial x}}\left(  \frac{\partial\rho
^{\mathrm{A}}}{\partial x}+\left(  \frac{\partial\rho^{\mathrm{A}}}{\partial
x}\rho^{\mathrm{B}}-{\frac{\partial\rho^{\mathrm{B}}}{\partial x}}%
\rho^{\mathrm{A}}\right)  \right) \nonumber\\
{\frac{\partial\rho^{\mathrm{B}}}{\partial t^{\prime}}}+{\frac{\partial
j^{\mathrm{B}}(\rho^{\mathrm{A}},\rho^{\mathrm{B}})}{\partial x}}  &
=\varepsilon{\frac{\partial}{\partial x}}\left(  \frac{\partial\rho
^{\mathrm{B}}}{\partial x}+\left(  \frac{\partial\rho^{\mathrm{B}}}{\partial
x}\rho^{\mathrm{A}}-{\frac{\partial\rho^{\mathrm{A}}}{\partial x}}%
\rho^{\mathrm{B}}\right)  \right) \label{hydro}\\
\varepsilon={\frac{a}{2}}\rightarrow0;\ \  &  {\frac{\partial}{\partial t}%
}=2\varepsilon{\frac{\partial}{\partial t^{\prime}}}\nonumber
\end{align}
For a finite system with open boundaries this is supplemented with the boundary 
conditions
\begin{equation}
\rho^{\mathrm{Z}}(0,t)=\rho_{L}^{\mathrm{Z}},\text{ \ }\rho^{\mathrm{Z}%
}(N,t)=\rho_{R}^{\mathrm{Z}}\text{\ , \ \ \ \ \ }Z=A,B\text{\ \ } \label{BC}%
\end{equation}
The system of equations (\ref{hydro}), after the substitutions $\rho
=1-\rho^{A}-\rho^{B},u=\rho^{B}-\rho^{A}$, becomes%
\begin{align}
{\frac{\partial\rho}{\partial t}+\frac{\partial}{\partial x}}\left(  \rho
u\right)   &  =\varepsilon{\frac{\partial^{2}\rho}{\partial x^{2}}%
}\label{Leroux}\\
{\frac{\partial u}{\partial t}+\frac{\partial}{\partial x}}\left(  \rho
+u^{2}\right)   &  =\varepsilon\frac{\partial}{\partial x}\left(
\frac{\partial u}{\partial x}+\left(  u\frac{\partial\rho}{\partial x}%
-\rho\frac{\partial u}{\partial x}\right)  \right)  ,\nonumber
\end{align}
which in its inviscid form ($\varepsilon=0$) has been derived rigorously for
our model defined on the infinite lattice \cite{FritzToth} and is known in 
the theory of conservation  law equations as the Leroux%
%TCIMACRO{\U{b4}}%
%BeginExpansion
\'{}%
%EndExpansion
s equation. It appears in many applications, further details about this
equation can be found in \cite{FritzToth,Serre2000}. Note that the viscosity
term is required for uniqueness of the solution of the hydrodynamic equation.
For an infinite system the choice of viscosity has some arbitraryness. However,
in the presence of the boundaries the exact form of the viscosity matrix
becomes important in systems with more than one conservation law
\cite{Mario_multiASEP}. The question how the macroscopic boundary condition (\ref{BC})
on the PDE
can be realized on microscopic lattice gas scale is addressed in the following
sections.

\section{Boundary reservoirs and projection measures}

In driven (nonequilibrium) systems, the steady state is governed not only by
local interactions in the bulk, but also by interaction with boundaries. To
define a boundary reservoir, we introduce a projection measure for a subsystem
of a larger system (defined on a spatial domain $\Omega$) by defining the following
marginal: a measure is a
\textit{projection measure } for a spatial domain $Q\subseteq\Omega$ if \textit{all
its correlations} are identical to those of the whole system $\Omega$\ inside
the domain $Q$.

In what follows we consider as domain $Q$ a finite system of $N$ sites as a
subsystem of an infinite system where $\Omega$ is the integer lattice $Z$. For
illustration we first consider $N=\infty$ (semi-infinite system with a left
boundary). In the case of a product measure, the projection measure is
especially simple: it remains a product measure for all sites in $Q$, i.e.,
homogeneous initial distribution with an average densities $\rho
_{\mathrm{bulk}}^{A},\rho_{\mathrm{bulk}}^{B}$ of $A$ and $B$ particles. In
vector notation \cite{Schu00} the projection measure for the semi-infinite system
is given by
\begin{equation}
|P(0)\rangle=%
%TCIMACRO{\dprod \limits_{k=1}^{\infty}}%
%BeginExpansion
{\displaystyle\prod\limits_{k=1}^{\infty}}
%EndExpansion
\otimes%
\begin{pmatrix}
\rho_{\mathrm{bulk}}^{A}\\
\rho_{\mathrm{bulk}}^{B}\\
1-\rho_{\mathrm{bulk}}^{B}-\rho_{\mathrm{bulk}}^{A}%
\end{pmatrix}
. \label{projection_half_infinite}%
\end{equation}

Similarly, for a finite subsystem of the length $N$ one has the projection measure%
\begin{equation}
|P\rangle=%
%TCIMACRO{\dprod \limits_{i=1}^{N}}%
%BeginExpansion
{\displaystyle\prod\limits_{i=1}^{N}}
%EndExpansion
\otimes%
\begin{pmatrix}
u\\
v\\
1-u-v
\end{pmatrix}
, \label{projection_finite}%
\end{equation}
where $u=\rho_{\mathrm{bulk}}^{A},v=\rho_{\mathrm{bulk}}^{B}$.

So far we have discussed only measures.
To make the connection to stochastic dynamics we assume that the particle
system defined on the infinite lattice has a product measure as its steady
state. In order to define dynamics for an open system with creation and
annihilation of particles corresponding to a boundary reservoir 
with densities $\rho_{L}%
^{A}=u,\rho_{L}^{B}=v$ for a semi-infinite system, we demand that the
projection measure (\ref{projection_half_infinite}) stays invariant for
$u=\rho_{\mathrm{bulk}}^{A},v=\rho_{\mathrm{bulk}}^{B}$. This is a natural
requirement, since one has a perfect match of the product state with the
boundaries. In the same way, we can define the right boundary reservoir for a
semi-infinite chain $k=-\infty,0$, and for a finite chain of the length $N$,
with the projection measure (\ref{projection_finite}), where  $\rho_{L}%
^{A}=\rho_{R}^{A}=u$, $\rho_{L}^{B}=\rho_{R}^{B}=v$. Some examples
illustrating more general situations (e.g., involving an Ising measure) can be
found in \cite{Gunter_Slava_Europhys,Peschel,TiborAntal}.

In general, there are many ways to obtain boundary rates which leave the
projection measure invariant. In order to calculate boundary rates for our
model we first require that particles are injected only at the edges (sites
$1,N$) of the lattice, corresponding to the nearest neighbor hopping in the
bulk of the chain. Furthermore
we note that, in particular, one requires the stationary flux in the
system with equal reservoir densities $\rho_{L}^{A}=\rho_{R}^{A}=u$, $\rho
_{L}^{B}=\rho_{R}^{B}=v$ to be equal to the stationary flux in an infinite
system with the densities $u,v$. The stationary flux is
\begin{align}
j^{A}(u,v)  &  =u(1-u+v)\label{jA}\\
j^{B}(u,v)  &  =-v(1-v+u). \label{jB}%
\end{align}
The different signs of $j^{A}$ and $j^{B}$ are due to the fact that
$A$-particles and $B$-particles hop in opposite directions. On the other hand,
the stationary flux of particles $A$ through the left boundary is
\begin{equation}
j_{L}^{A}=\langle0\rangle\Gamma_{A0}^{L}+\langle B\rangle\Gamma_{AB}^{L}
\label{jAL}%
\end{equation}
where $\Gamma_{XY}^{L}$ is the rate of exchange of particle of the sort $X$
(from the left reservoir) with the particle of sort $Y$ at the first site and
$\langle Y\rangle$ is the stationary expectation value to find a particle $Y$
at the first site. In the stationary state, the latter is equal to the density
(see (\ref{projection_finite})): $\langle B\rangle=v$, \ $\langle
0\rangle=1-u-v$. Inserting these expressions in (\ref{jAL}), we get%
\[
j_{L}^{A}=(1-u-v)\Gamma_{A0}^{L}+v\Gamma_{AB}^{L}%
\]
Repeating the arguments for the flux of particles $B$ through the left
boundary, we obtain%
\[
j_{L}^{B}=-\langle B\rangle\Gamma_{0B}^{L}-\langle B\rangle\Gamma_{AB}%
^{L}=-v(\Gamma_{0B}^{L}+\Gamma_{AB}^{L})
\]
We require the above expressions to be equal to the stationary fluxes
(\ref{jA}),(\ref{jB}).This leaves one parameter undefined, since there are $3$
unknown boundary rates $\Gamma$ and two relations. To get an additional
relation, consider another correlation, e.g. the probability to find particles
$A,B$ on the first two sites, $\langle AB\rangle$. The time evolution of this
correlation is
\[
\frac{\partial}{\partial t}\langle AB\rangle=\Gamma_{A0}^{L}\langle
0B\rangle+\Gamma_{0B}\langle A0B\rangle+\Gamma_{AB}\langle AAB\rangle
-\Gamma_{AB}\langle AB\rangle.
\]
In the stationary state, left hand side vanishes, and the correlations are
given by (\ref{projection_finite}) $\langle0B\rangle=(1-u-v)v,$ $\langle
AB\rangle=uv,$ $\langle AAB\rangle=u^{2}v,$ $\langle A0B\rangle=u(1-u-v)v$.
Substituting the bulk rates $\Gamma_{XY}$ from (\ref{rates112}), we get
$\Gamma_{A0}^{L}=u$, which, together with the relations $j_{L}^{A}=j^{A},$
$j_{L}^{B}=j^{B}$ fixes all boundary rates to be
\begin{equation}
\Gamma_{A0}^{L}(u,v)=u;\text{ \ \ }\Gamma_{AB}^{L}(u,v)=2u;\text{ \ }%
\Gamma_{0B}^{L}(u,v)=(1-u-v).\text{ \ \ \ } \label{solL}%
\end{equation}
It is interesting to note that the rates (\ref{solL}) can be obtained from
another simple argument. In a boundary reservoir with the densities $\rho
_{L}^{A}=u$, $\rho_{L}^{B}=v$, the probability to find a particle $A,B$ and a
hole $0$ is $p(A)=u$, $p(B)=v$, $p(0)=1-u-v$. We assume the rates of hopping
of the particles from the reservoir to the site $1$ of the system to be given
by the bulk rates (\ref{rates112}) multiplied by corresponding probabilities,
i.e.,%
\begin{equation}
\Gamma_{A0}^{L}=p(A)\Gamma_{A0}\text{, \ }\Gamma_{0B}^{L}=p(0)\Gamma
_{0B}\text{, \ \ }\Gamma_{AB}^{L}=p(A)\Gamma_{AB}.\text{\ }
\label{reservoir_L_rates}%
\end{equation}
This yields (\ref{solL}).

Analogously, on the right boundary, the rates $\Gamma_{XY}^{R}$ of exchange of
particle $X$ at site $N$ with the particle $Y$ (from the right reservoir) is
\begin{equation}
\Gamma_{A0}^{R}=p(0)\Gamma_{A0}\text{, \ }\Gamma_{0B}^{R}=p(B)\Gamma
_{0B}\text{, \ \ }\Gamma_{AB}^{R}=p(B)\Gamma_{AB} \label{reservoir_R_rates}%
\end{equation}
or
\begin{equation}
\Gamma_{A0}^{R}(u,v)=(1-u-v),\text{ \ \ }\Gamma_{0B}^{R}(u,v)=v,\text{
\ \ \ }\Gamma_{AB}^{R}(u,v)=2v. \label{solR}%
\end{equation}
One verifies straightforwardly that for the choice of the rates (\ref{solL}),
(\ref{solR}) the projection measure (\ref{projection_finite}) is stationary.

For the boundary reservoirs with arbitrary densities $\rho_{R}^{A},\rho
_{R}^{B}$ on the right and $\rho_{L}^{A},\rho_{L}^{B}$ on the left we shall
use the boundary rates given by $\Gamma_{XY}^{R}(\rho_{R}^{A},\rho_{R}^{B})$
and $\Gamma_{XY}^{L}(\rho_{L}^{A},\rho_{L}^{B})$ respectively. We shall call
such boundary rates PDE-friendly, since, as we show in the next section, they
allow for an adequate hydrodynamic description for a finite open system.

\section{\bigskip Infinite reflections of shock fronts from the boundaries.
Absence of spontaneous symmetry breaking.}

It is an open problem how to obtain a hydrodynamic description of a two
species system with open boundaries. A step towards such a description has
been made in \cite{reflections_JPA}. It has been demonstrated that the
evolution to the stationary state can proceed through an infinite sequence of
reflections of a domain wall with the boundaries of the system. It is
interesting to study the reflections for the system (\ref{rates112}) to see if
it provides another example where this scenario applies. Note that the model
we investigate here and the one investigated in \cite{reflections_JPA} have
topological differences: in the latter, particle dynamics on two parallel
chains was considered, and particles on both chains hopped in the same
direction. Instead, in the presently considered model (\ref{rates112}),
particles occupy one chain and hop in opposite directions.

\subsection{Reflection maps}

To pose a reflection problem, take a half-infinite chain with only one, e.g.,
left boundary, corresponding to left boundary reservoir densities $\rho
_{L}^{A}=u,\rho_{L}^{B}=v$, and choose an \textit{initial state} to be a
projection measure of a stationary state for an infinite system onto the
positive half-infinite chain (\ref{projection_half_infinite}), i.e.,
homogeneous initial distribution with an average densities $\rho
_{\mathrm{bulk}}^{A},\rho_{\mathrm{bulk}}^{B}$ of $A$ and $B$ particles. We
consider bulk rates (\ref{rates112}) and corresponding boundary rates
(\ref{solL}).

\begin{figure}
%[ptb]
\begin{center}
\includegraphics[
height=3.5405in,
width=5.0548in
]%
{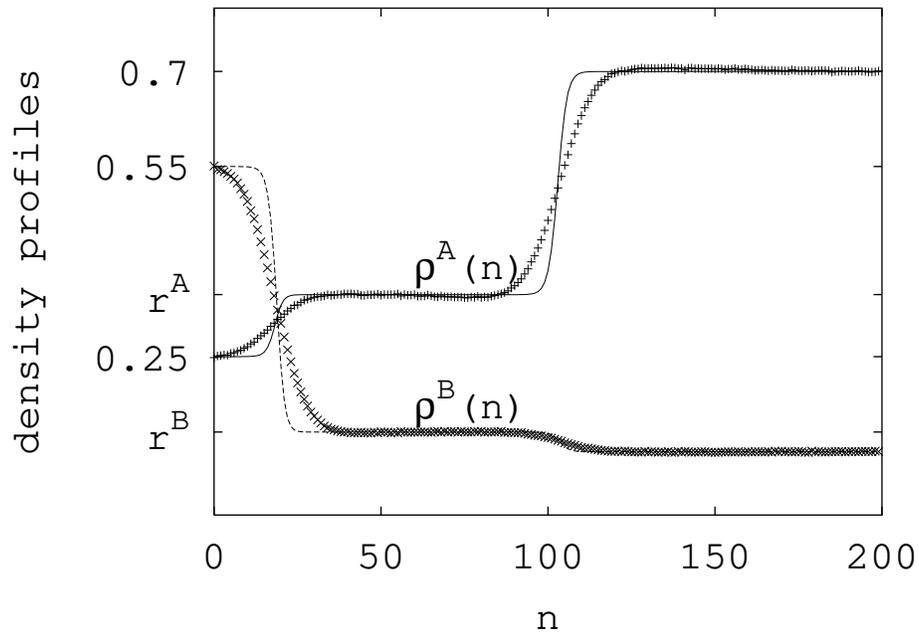}%
\caption{Domain wall reflection from the left boundary, as given by Monte
Carlo simulations. Particles were initially distributed randomly without
correlations with constant densities $\rho^{A}=0.7,\rho^{B}=0.1$, fitting the
right boundary. The average profiles after $300$ Monte Carlo steps are shown,
and compared with the hydrodynamic limit (\ref{hydro}) evolution (continuous
curves). Averaging over $5\ast10^{5}$ histories is made. The smoothing of the
Monte-Carlo density profile is due to fluctuations in the shock position which
is scaled out in the hydrodynamic description. }%
\label{fig_MC_Lrefl}%
\end{center}
\end{figure}

If the bulk densities coincide with the left boundary densities $\rho
_{\mathrm{bulk}}^{A}=u,\rho_{\mathrm{bulk}}^{B}=v$, the initial distribution
(\ref{projection_half_infinite}) stays invariant because of the perfect match
of the initial state with the boundaries. Otherwise, there is a mismatch at
the boundary which has to be resolved. In this case, Monte Carlo simulations
lead to one of the following scenarios: (a) a thin boundary layer develops
interpolating between the bulk densities $\rho_{\mathrm{bulk}}^{A,B}$ and
boundary densities $u,v$, and stays always attached to the left boundary, (b)
a shock wave of the densities $r^{A},r^{B}$ develops and propagates to the
right, (c) a rarefaction wave with the density $r^{A},r^{B}$ forms and spreads
to the right. The scenario (b) is demonstrated on Fig.\ref{fig_MC_Lrefl},
showing Monte-Carlo evolution of a state (\ref{projection_half_infinite}) with
$\rho_{\mathrm{bulk}}^{A,B}=(0.1,0.7),$ \ $u,v=0.25,0.55$ after $t=300$ Monte
Carlo steps. Like in \cite{reflections_JPA}, we mimic the half-infinite chain
with a finite chain with the right boundary conditions matching perfectly the
bulk densities. As a result, after time $t$ the density profile, while staying
unchanged on the right boundary (see Fig.\ref{fig_MC_Lrefl}), at the left
boundary develops the shock with the densities $r^{A}\approx0.348,r^{B}%
\approx0.131$ spreading to the right. Due to particle number conservation in
the bulk, a $Z$-component of the shock interface moves with the velocity
\begin{equation}
V^{Z}={\frac{j_{\mathrm{bulk}}^{Z}-j_{\mathrm{r}}^{Z}}{\rho_{\mathrm{bulk}%
}^{Z}-r^{Z}}}\ , \label{v_shock}%
\end{equation}
where we used the short notations $j_{\mathrm{bulk}}^{Z}=j^{Z}(\rho
_{\mathrm{bulk}}^{A},\rho_{\mathrm{bulk}}^{B})$, $j_{\mathrm{r}}^{Z}%
=j^{Z}(r^{A},r^{B})$. Since there is an interaction between $A$ and $B$
species, the velocities must coincide in both components: $V^{A}=V^{B}$, or
\begin{equation}
{\frac{j_{\mathrm{bulk}}^{A}-j_{\mathrm{r}}^{A}}{\rho_{\mathrm{bulk}}%
^{A}-r^{A}}}={\frac{j_{\mathrm{bulk}}^{B}-j_{\mathrm{r}}^{B}}{\rho
_{\mathrm{bulk}}^{B}-r^{B}}}, \label{vAvB}%
\end{equation}
defining implicitly the allowed location of the points $r^{A},r^{B}$. In our
case, the above relation becomes (see \ref{fluxA},\ref{fluxB}):%
\begin{equation}
\frac{\rho_{\mathrm{bulk}}^{A}(1-\rho_{\mathrm{bulk}}^{A}+\rho_{\mathrm{bulk}%
}^{B})-r^{A}(1-r^{A}+r^{B})}{\rho_{\mathrm{bulk}}^{A}-r^{A}}=\frac
{-\rho_{\mathrm{bulk}}^{B}(1-\rho_{\mathrm{bulk}}^{B}+\rho_{\mathrm{bulk}}%
^{A})+r^{B}(1-r^{B}+r^{A})}{\rho_{\mathrm{bulk}}^{B}-r^{B}} \label{vAvB_our}%
\end{equation}
Here we implicitly assume the fluxes of the reflected wave to be given by
stationary fluxes with the densities $r^{A},r^{B}$. This supposition is
justified because for random initial conditions used here the system
away from the shock is locally stationary.
%the time needed to establish local stationarity is much
%shorter than the time to establish global stationarity (for finite systems,
%the latter is at least of the order of the time needed for a shock to reach
%the other boundary).

\begin{figure}
%[ptb]
\begin{center}
\includegraphics[
height=3.5405in,
width=5.0548in
]%
{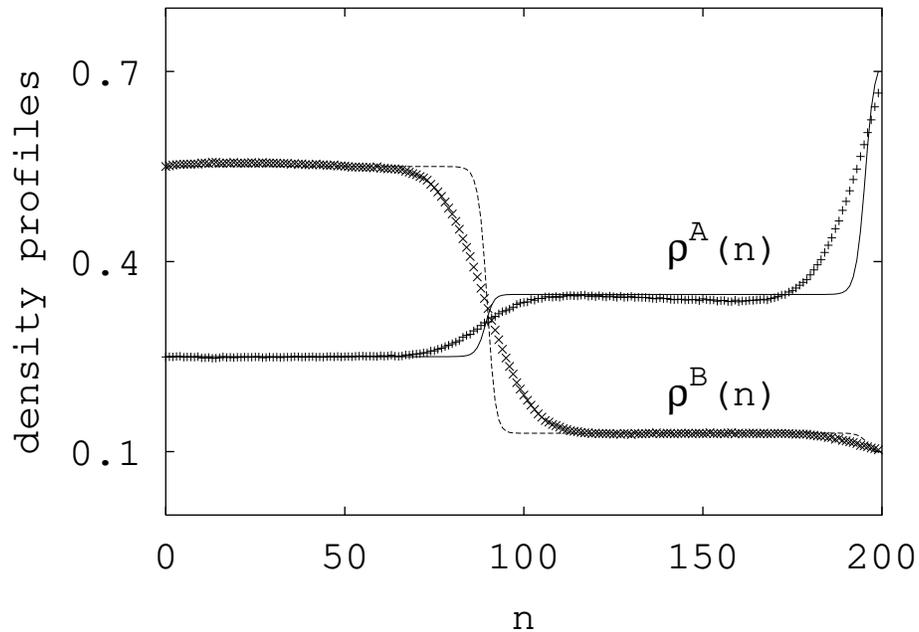}%
\caption{Domain wall reflection from the right boundary. Particles were
initially distributed randomly without correlations with constant densities
$\rho^{A}=0.25,\rho^{B}=0.55$ fitting the left boundary density. Average
density profiles after $t=300$ MCS steps are shown. The boundary densities are
as on Fig. \ref{fig_MC_Lrefl}. Lines show the hydrodynamic evolution
(\ref{hydro}) for the two particle species. The smoothing of the Monte-Carlo
density profile is due to fluctuations in the shock position which is scaled
out in the hydrodynamic description.}%
\label{fig_MC_Rrefl}%
\end{center}
\end{figure}

Analogously one can consider the right boundary reflection, taking a
half-infinite chain with right boundary only and initial product measure state
(\ref{projection_half_infinite}) where the product goes from $-\infty$ to $L$.
An example of a right boundary reflection is shown on Fig.\ref{fig_MC_Rrefl}.
Again, the shock wave resolving the mismatch at the boundary appears and
spreads to the left. For the same reasons as were given above for left
boundary reflection, the densities of reflected waves $r^{A},r^{B}$ satisfy
relation (\ref{vAvB_our}). Solutions of (\ref{vAvB_our}) are two families of
straight lines
\begin{equation}
r^{B}-\rho_{\mathrm{bulk}}^{B}=(r^{A}-\rho_{\mathrm{bulk}}^{A})\gamma_{+}%
(\rho_{\mathrm{bulk}}^{A},\rho_{\mathrm{bulk}}^{B}),\text{ \ \ \ \ \ \ }%
\gamma_{+}(s,t)=\frac{1}{2s}\left(  -2+s+t+\sqrt{(s+t-2)^{2}-4st}\right)
\text{.} \label{sol_L}%
\end{equation}%
\begin{equation}
r^{B}-\rho_{\mathrm{bulk}}^{B}=(r^{A}-\rho_{\mathrm{bulk}}^{A})\gamma_{-}%
(\rho_{\mathrm{bulk}}^{A},\rho_{\mathrm{bulk}}^{B}),\text{ \ \ \ \ \ \ }%
\gamma_{-}(s,t)=\frac{1}{2s}\left(  -2+s+t-\sqrt{(s+t-2)^{2}-4st}\right)
\text{.} \label{sol_R}%
\end{equation}
The solutions with bigger $\gamma$ corresponds to the left reflection, and
another to the right reflection. Note that $\gamma_{\pm}(s,t)$ are real in the
physical domain $0\leq s+t\leq1$. The meaning of the above equations is
twofold. One of them was given above, that is, for any given $\rho
_{\mathrm{bulk}}^{A},\rho_{\mathrm{bulk}}^{B}$, the reflected waves must have
densities satisfying (\ref{sol_L}),(\ref{sol_R}). Another, alternative, is:
given the reflected wave densities $r^{B}$ and $r^{A}$, the initial bulk
densities ( which have led to those reflected waves), must satisfy the
equation equivalent to (\ref{sol_L},\ref{sol_R}).
\begin{equation}
\rho_{\mathrm{bulk}}^{B}-r^{B}=(\rho_{\mathrm{bulk}}^{A}-r^{A})\gamma_{\pm
}(r^{A},r^{B}), \label{sol_for_bulk_densities}%
\end{equation}
where $\gamma_{\pm}(s,t)$ are defined in (\ref{sol_L},\ref{sol_R}). The
equation above is a solution of (\ref{vAvB_our}), because the latter is
completely symmetric with respect to the exchange $r^{Z}\leftrightarrow$
$\rho_{\mathrm{bulk}}^{Z}$. On the basis of Monte-Carlo calculations and
investigation of the hydrodynamic limit equations the following observation
can be made:

\textit{The collection of all possible densities }$r^{A},r^{B}$\textit{ of the
waves, resulting from the left boundary reflection (with boundary densities
}$u,v$\textit{) with whatever initial conditions }$\rho_{\mathrm{bulk}}%
^{A},\rho_{\mathrm{bulk}}^{B}$\textit{, constitute some curve } $\mathcal{L}%
(u,v)$\textit{, completely parametrized by this point }$u,v$\textit{, and
containing this point. All initial bulk densities satisfying
(\ref{sol_for_bulk_densities}) for certain point }$r^{A},r^{B}$\textit{ of
}$\mathcal{L}$\textit{, result in the reflected wave with the densities
}$r^{A},r^{B}$\textit{. The same is true for right boundary reflection, with
the curve }$\mathcal{L}$ replaced by $\mathcal{R}$.

\begin{figure}
%[ptb]
\begin{center}
\includegraphics[
height=3.5405in,
width=5.0548in
]%
{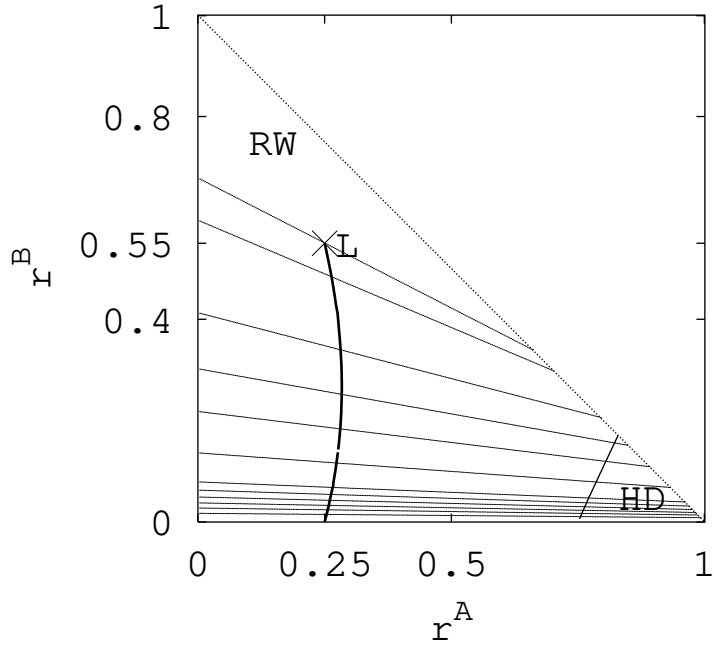}%
\caption{Left reflection map for the left boundary densities $0.25,0.55$ of
$A$ and $B$ particles respectively. The crossmark labelled $L$ marks the point
of the left boundary density. The curve containing this point $L$ is a
location of all possible densities $r^{A},r^{B}$ of the \ reflected waves,
except at the region HD. The straight lines show initial bulk densities with
the same outcome of a reflection (point of intersection of the straight line
with the curve). In the HD region the initial wave stays glued to the left
boundary. In the RW region (the upper triangle) the reflection results in a
rarefaction wave (scaling as $(x/t)$ with time), converging to point $L$. The
dotted line $r^{A}+r^{B}=1$ shows the boundaries of the physical region.}%
\label{fig_L}%
\end{center}
\end{figure}

\begin{figure}
%[ptbptb]
\begin{center}
\includegraphics[
height=3.5405in,
width=5.0548in
]%
{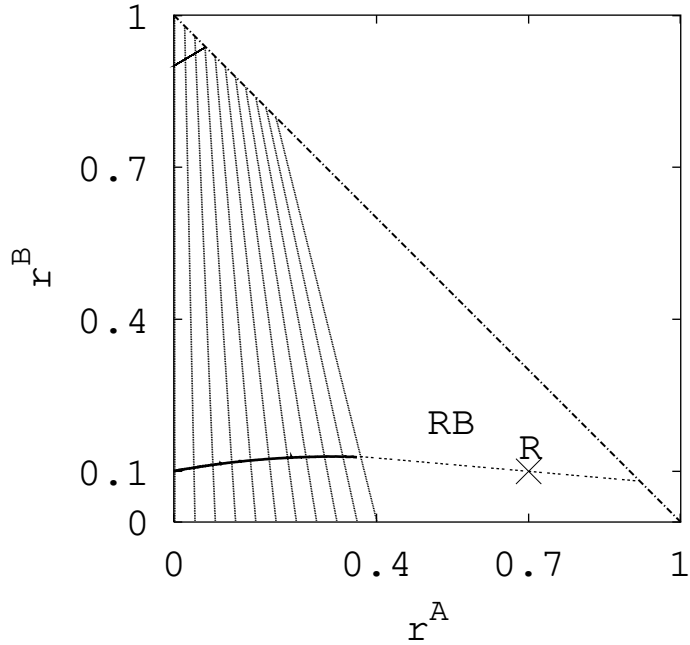}%
\caption{Right reflection map for the right boundary density $0.1,0.7$ of
$A$,$B$ particles. The crossmark labelled $R$ marks the point of the right
boundary density. The thick curve is a location of all possible densities
$r^{A},r^{B}$ of the \ reflected waves, except at the upper left triangle and
in region RB. Straight lines show initial bulk densities with the same outcome
of a reflection (point of intersection of the line with the thick curve). In
the upper corner the initial wave stays glued to the right boundary. In the
region RB an intermediate plateau appears, with the densities corresponding to
the broken line passing through point $R$. The final result of this reflection
is a wave matching the right boundary $R$. The dotted line $r^{A}+r^{B}=1$
shows the boundaries of the physical region.}%
\label{fig_R}%
\end{center}
\end{figure}

The mapping from arbitrary $\rho_{\mathrm{bulk}}^{A},\rho_{\mathrm{bulk}}^{B}$
to resulting $r^{A},r^{B}$ is described by means of reflection maps introduced
in \cite{reflections_JPA}. Typical examples of left and right reflection maps
are shown on Figs. \ref{fig_R} and \ref{fig_L}.

\begin{figure}
%[ptb]
\begin{center}
\includegraphics[
height=3.5405in,
width=5.0548in
]%
{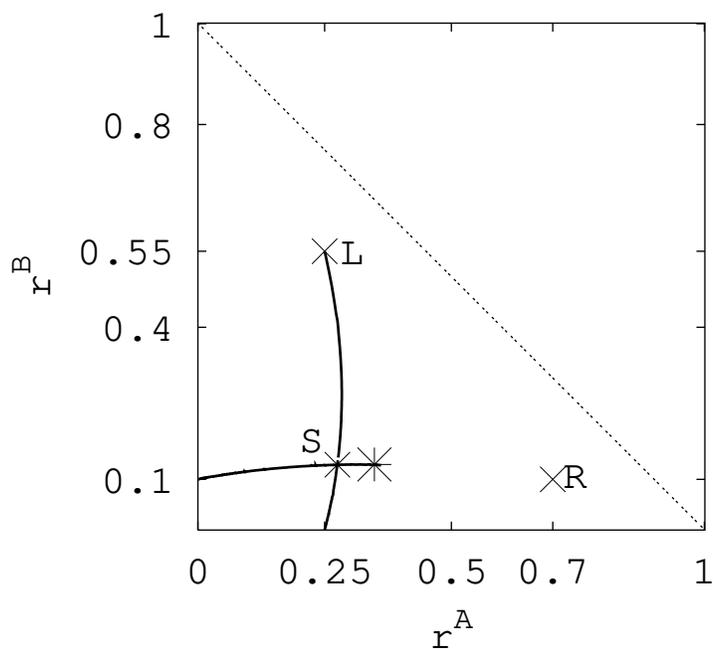}%
\caption{ Curves of the right and left boundary reflection combined together.
The intersection $S$, is the stationary state density achieved through the
infinite number of reflections. The crossmarks labelled by $R$ and $L$
indicate the left and right boundary densities, and the cross at the
intersection the stationary density obtained by Monte Carlo simulation of the
system of $300$ sites. The system was equilibrated for = $4\ast10^{5}$ Monte
Carlo Steps (MCS), after which the averaging over again $4\ast10^{5}$ MCS and
$10$ different histories was done. The big star to the righthand side of $S$
marks the result of a single reflection for an initial profile fitting the left
boundary density point $L$ (from Monte Carlo calculations).}%
\label{fig_RL}%
\end{center}
\end{figure}

\begin{figure}
%[ptbptb]
\begin{center}
\includegraphics[
height=4.0577in,
width=4.1303in
]%
{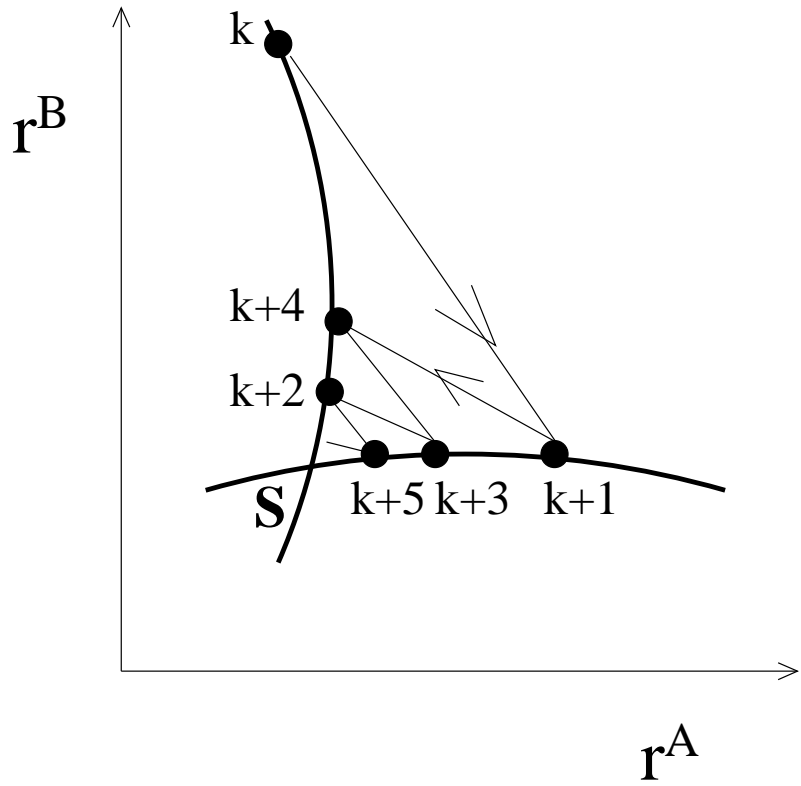}%
\caption{ Approach to the stationary point $S$ through an infinite sequence of
reflections. The filled circles show the location of subsequent domain wall
densities $r^{A},r^{B}$. The circles $k,k+2,k+4$ correspond to the result of
the left reflection, while the other circles correspond to the right
reflection.}%
\label{fig_AB0_enlarge}%
\end{center}
\end{figure}

Each of the reflection maps of the type shown in Figs. \ref{fig_L} and
\ref{fig_R} is parametrized \ by a single point $u,v$ (which has a meaning of
the left/right boundary density for left/right map). Each point $0\leq
u+v\leq1$ corresponds to a different map. Classification of all different maps
requires separate investigation. Here we shall find out what happens if we fix
the right and the left boundary corresponding to the reflection maps
Fig.~\ref{fig_R} and Fig.~\ref{fig_L}. Let us take an initial homogeneous
state with the densities, fitting the density of the left reservoir,
$\rho_{\mathrm{bulk}}^{A},\rho_{\mathrm{bulk}}^{B}=\rho_{L}^{A},\rho_{L}^{B}$.
As the result of the first interaction with the right boundary, a reflected
wave forms, with the densities $r_{0}^{A}$, $r_{0}^{B}$ \ given by the
intersection of the line (\ref{sol_R}) with the curve $\mathcal{R}$ of
Fig.~\ref{fig_R}. This reflected wave hits the left boundary then, resulting
in the next reflected wave with the densities $r_{1}^{A}$ ,$r_{1}^{B}$. The
left reflection is controlled by the curve $\mathcal{L}$ at Fig.~\ref{fig_L},
therefore the densities of the reflected wave $r_{1}^{A}$, $r_{1}^{B}$ are
given by the intersection of the line (\ref{sol_L}), where new $\rho
_{\mathrm{bulk}}^{A},\rho_{\mathrm{bulk}}^{B}=r_{0}^{A}$, $r_{0}^{B}$, with
the curve $\mathcal{L}$. Since the curves $\mathcal{R}$,$\mathcal{L}$ do not
coincide with any of the lines defined by (\ref{solL}), (\ref{solR}), this
process continues forever, though converging to the stationary state
$\mathcal{S}$, the point of the intersection of $\mathcal{L}$ with
$\mathcal{R}$, see Fig.~\ref{fig_RL}. After $2k-1$ reflections, the wave with
the densities $r_{2k-1}^{A}$, $r_{2k-1}^{B}$ will hit the right boundary,
producing the reflected wave of the densities $r_{2k}^{A}$, $r_{2k}^{B}$,
corresponding to the intersection of the line (\ref{sol_R}) ( with
$\rho_{\mathrm{bulk}}^{A},\rho_{\mathrm{bulk}}^{B}=r_{2k-1}^{A}$,
$r_{2k-1}^{B}$) with the curve $\mathcal{R}$. The latter wave, hitting the
left boundary, produces the next reflected wave, and so on. This process is
depicted schematically on Fig.~\ref{fig_AB0_enlarge}. Several remarks are in order.

\begin{itemize}
\item The densities $\ \{r_{k}^{A}$, $r_{k}^{B}\}_{k=0}^{\infty}$ of the
reflected waves constitute a converging sequence. Convergence is exponential,
$\delta u_{n+2k}=e^{-\kappa k}\delta u_{n}$, for $n\rightarrow\infty$ where we
denote by $\delta u_{n}$ the deviation from the stationary density at the
$n$-th step. $e^{-\kappa}<1$ is a constant depending on tangential derivatives
$\alpha,\beta,\alpha_{1},\beta_{1}$ of the curves $\mathcal{R}$,$\mathcal{L}$
and the characteristic curves(\ref{solL}), (\ref{solR}), respectively, at the
stationary point $\mathcal{S}$, $\ e^{-\kappa}=(1-\frac{tg(\alpha_{1}%
)}{tg(\alpha)})(1-\frac{tg(\beta_{1})}{tg(\beta)})/((1+\frac{tg(\alpha_{1}%
)}{tg(\beta)})(1-\frac{tg(\beta_{1})}{tg(\alpha)}))$. Note that if \ at least
one of the characteristic derivatives happens to coincide with $\alpha
_{1},\beta_{1} (\alpha=\alpha_{1}$ or $\beta=\beta_{1}$)  then the sequence
converges in one step.

\item The velocities of reflected waves for large $k$ converge to the finite
characteristic velocities computed at the stationary point $\mathcal{S}$, as
eigenvalues of the flux Jacobian, see \cite{GunterSlava_StatPhys}.

\item The stationary densities are not reached at any finite time. Strictly
speaking, the stationary density is reached exponentially in time with
characteristic time proportional to the length of the system $N$. On the
contrary, in one-species models like the ASEP, the corresponding
characteristic time is of order of 1/(rate of hopping).
\end{itemize}

\begin{figure}
%[ptbptbptb]
\begin{center}
\includegraphics[
height=3.5405in,
width=5.0548in
]%
{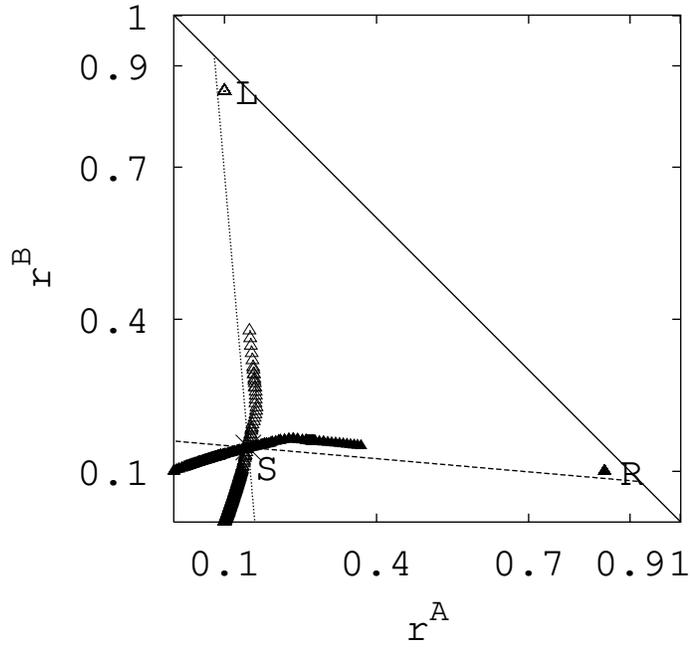}%
\caption{ Curves of the right (filled triangles) and left (empty triangles)
boundary reflection for the symmetric setting
(\ref{symmetric_boundary_conditions}), as given by numerical integration of
the hydrodynamic equations (\ref{hydro}). Triangles labelled $R$ and $L$
indicate left and right boundary densities, and the point $S$ the stationary
density, achieved through an infinite series of reflections. The cross at the
intersection point $S$ marks the stationary density obtained by Monte Carlo
simulation of the system of $300$ sites. The broken and dotted lines
intersecting at $S$ show characteristic curves at the point $S$ (see
(\ref{sol_L}),(\ref{sol_R})). }%
\label{fig_symm_Map}%
\end{center}
\end{figure}

We conclude that for the system (\ref{rates112}) with PDE-friendly boundary
rates $\Gamma_{XY}^{L}(\rho_{L}^{A},\rho_{L}^{B}),\Gamma_{XY}^{R}(\rho_{R}%
^{A},\rho_{R}^{B})$, given by (\ref{solL}), (\ref{solR}), the convergence to a
stationary state through an infinite number of reflections is a generic
feature of the dynamics. This is confirmed by the good agreement of the
Monte-Carlo simulation of the lattice model with the numerical integration of
hydrodynamic PDE's (\ref{hydro}), see Figs.\ref{fig_MC_Lrefl},
\ref{fig_MC_Rrefl}, \ref{fig_RL}, \ref{fig_symm_Map}. Without the PDE-friendly
boundary rates already the result of the first reflection would be unpredictable.

\subsection{Symmetric boundary conditions}

In the preceeding subsection we gave an example for the time evolution of the
system in case of generic boundary conditions. Here we consider a situation
when the boundary rates possess the symmetry with respect to a simultaneous
right-left reflection and an exchange of $A$ particles with $B$ particles,
i.e., the same symmetry as in the similar models where spontaneous symmetry
breaking has been observed \cite{Mukamel95,MukamelToyModel}. This
amounts to the following restriction of the boundary reservoir densities:%
\begin{equation}
\rho_{R}^{A}=\rho_{L}^{B},\text{ \ \ }\rho_{R}^{B}=\rho_{L}^{A}.
\label{symmetric_boundary_conditions}%
\end{equation}
Analysis of reflection maps for our PDE-friendly boundary rates (\ref{solL}%
,\ref{solR}) allows us to conclude, that there is no spontaneous symmetry
breaking --- the stationary state is always symmetric (i.e., corresponds to
equal densities of $A$- and $B$-particles). A typical reflection map for the
case (\ref{symmetric_boundary_conditions}) is presented in
Fig.\ref{fig_symm_Map}. The reason why there is spontaneous symmetry breaking
in the bridge model \cite{Mukamel95} will be given in the next section.

\section{Why the hydrodynamic limit fails for the bridge model}

The bridge model is defined on a finite chain of the length $N$, where the
processes in the bulk with rates
\begin{equation}
\Gamma_{A0}=\Gamma_{0B}=1,\Gamma_{AB}=q,\text{ \ \ \ }\Gamma_{0A}=\Gamma
_{B0}=\Gamma_{BA}=0\text{\ } \label{bridge_bulk_rates}%
\end{equation}
are complemented with the processes at the boundaries. On the left boundary
site
\begin{equation}
\Gamma_{A0}^{L}=\alpha=\alpha\Gamma_{A0}\text{, \ }\Gamma_{0B}^{L}=\beta
=\beta\Gamma_{0B}\text{, \ \ }\Gamma_{AB}^{L}=0
\label{bridge_boundary_rates_L}%
\end{equation}
and on the right boundary site,%
\begin{equation}
\Gamma_{0B}^{R}=\alpha=\alpha\Gamma_{0B}\text{, \ }\Gamma_{A0}^{R}=\beta
=\beta\Gamma_{A0}\text{, \ \ }\Gamma_{AB}^{R}=0.
\label{bridge_boundary_rates_R}%
\end{equation}
The model is invariant with respect to simultaneous left-right and $A$ -$B$
interchange. Nevertheless, for certain range of parameters, typically for
large $\alpha$ and small $\beta$, the state of the system is characterized by
a phase where the average densities of particles are not symmetric: the
symmetry between $A$ and $B$ is broken spontaneously, and it takes a system a
very long time (which grows exponentially with the system size) to get from
the state with a prevalence of $A$-particles to the state with a  prevalence
of $B$-particles. The exchange rate $q$ is not crucial for the occurence of
this intriguing phenomenon.

(a) Let us consider $q=2$ and try to see if one can associate a reservoir
density as described above to the rates (\ref{bridge_boundary_rates_L}),
(\ref{bridge_boundary_rates_R}). In terms of the probabilities
$p(A),p(B),p(0)$ to find a particle of sort $A,B$ and $0$ respectively in the
reservoir, the left reservoir rates are given in (\ref{reservoir_L_rates}).
Comparing (\ref{bridge_boundary_rates_L}) with (\ref{reservoir_L_rates}), we
obtain relations%
\[
p(A)=\alpha,p(0)=\beta,p(A)=0,
\]
incompatible between themselves.

(b) Let us nevertheless try to find\textit{ effective }boundary densities for
the rates (\ref{bridge_boundary_rates_L}, \ref{bridge_boundary_rates_R}) for
general $q$. It is instructive to analyze the case when both injection and
extraction are small $\alpha,\beta\ll1$ and $\alpha<\beta$. Then, the typical
time to inject a particle is bigger than the typical time to extract a
particle $\tau_{in}=\alpha^{-1}>\beta^{-1}=\tau_{out},$ and both $\tau
_{in},\tau_{out}\gg1.$ In this case, one can associate boundary densities as
it is done in the usual asymmetric exclusion process (ASEP): $\rho_{L}%
^{A}=\alpha,\rho_{L}^{B}=1-\beta$, and $\rho_{R}^{A}=1-\beta,\rho_{R}%
^{B}=\alpha$. Correspondingly, the density of holes in the boundary reservoirs
is $1-\rho^{A}-\rho^{B}$, yielding%
\begin{equation}
\rho_{R}^{0}=\rho_{L}^{0}=\beta-\alpha. \label{density of holes}%
\end{equation}
We see that for $\alpha>\beta$ the density of holes (\ref{density of holes})
becomes\textit{ negative}, which signals a breakdown of the reservoir picture
across the point $\alpha=\beta$ when $\alpha,\beta\ll1$. And this point lies
exactly at the phase transition line to the symmetry broken phase
\cite{MukamelToyModel}. Thus, we have shown that for parameters where symmetry
breaking phase is observed, the hydrodynamic description, using boundary
reservoirs, fails. A natural interpretation of the rates
(\ref{bridge_boundary_rates_L}), (\ref{bridge_boundary_rates_R}) would be that
they effectively correspond to impurities, placed at the left and at the right
boundary of the system, and preventing $A-B$ exchange. It is solely due to
these obstacles, that spontaneous symmetry breaking actually happens. An
effective hydrodynamic description may still be valid in the symmetric phase.

The above considerations are valid for (a) $q=2$, $\alpha,\beta$ arbitrary and
(b) $q$ arbitrary and $\alpha,\beta$ small. However, we believe that a similar
description is valid for any $q,\alpha,\beta$. Since for $q=2$ as well as for
$q=1$ there are no correlations between particles and vacancies, a perfect
hydrodynamic description of the bridge model will be achieved according to
the result of the previous section, if exchange at
the boundaries between $A,B$ happens \textit{on the same footing as in the
bulk}, see the discussion below (\ref{solL}), which amounts to choosing
\begin{equation}
\Gamma_{AB}^{L}=\Gamma_{AB}^{R}=q\alpha,\text{ \ \ \ }%
q=1,2\label{bridge_model_friendly}%
\end{equation}
in (\ref{bridge_boundary_rates_L}), (\ref{bridge_boundary_rates_R}). For the
above choice, a hydrodynamic description with fixed boundary densities 
is valid for all $\alpha,\beta$
leading to the complete disappearance of the symmetry broken phase from the
phase diagram. For instance, for $q=1$, the $A$-particle cannot distiguish
between $B$-particle and a hole, both in the bulk and in the boundaries (and
the same is valid for $B$-particles), so that the system effectively separates
into two one-species problem ASEP, with injection rate $\alpha$ and extraction
rate $\alpha+\beta$. Invoking the known solution for ASEP phase diagram
\cite{ASEP1},\cite{ASEP}, we find that the exact stationary state density of a
PDE-friendly bridge model with $q=1$ is
\begin{align*}
\rho_{stat}^{A} &  =\rho_{stat}^{B}=\alpha\text{, \ if }\alpha<\frac{1}{2}\\
\rho_{stat}^{A} &  =\rho_{stat}^{B}=\frac{1}{2}\text{, \ if }\alpha\geq
\frac{1}{2}.
\end{align*}
Note also that the stationary state in this case, as opposed to the $q=2$ case
considered earlier, is reached after \textit{ single} interaction of a shock
wave, or a rarefaction wave, with the boundary, like in the ASEP
\cite{reflections_JPA}.

\section{Conclusions}

We have investigated the ABO model and, defining projection measures,
introduced boundary conditions compatible with a hydrodynamic description. We
have studied the system for boundary rates which are either symmetric or
nonsymmetric with respect to left-right reflection and exchange of particle
species. Investigating the reflection of shocks at the boundaries, we conclude
that the system with symmetric PDE-friendly boundary conditions has a
stationary symmetric phase in all parameter space. The relaxation to the
stationary state proceeds by an infinite sequence of reflections with the
boundaries, which can be described using reflection maps. Hydrodynamic limit
equations are constructed and studied, yielding results consistent with the
stochastic dynamics. For the bridge model, which is a special case of the ABO
model, we showed that at the phase transition to the spontaneously broken
phase the hydrodynamic description fails. While our discussion was in the
framework of the ABO model, we believe that similar results are valid for
other driven diffusive systems with two conservation laws, such as two-lane
models \cite{Peschel,GunterSlava_StatPhys,Lahi00,Das01} or bricklayer models 
with nonconserved internal degrees of freedom
\cite{Valk03}.

We conclude that any choice of
boundary rates which does not correspond to certain densities of boundary
reservoirs is equivalent to placing an impurity at the respective boundaries.
This entails extra complexity and rate-dependent nonuniversal behaviour.
Sometimes the impurity thus introduced leads only to a local disturbance and
correspondingly to a redefinition of boundary densities. In other cases,
however, the impurity effect is highly nonlocal and nontrivial. This last case
is very pronounced in the case of spontaneous symmetry breaking considered in
\cite{Mukamel95}. It would be interesting to see if for any other choice
of the $A-B$ boundary exchange rates $0\leq\Gamma_{AB}^{L}=\Gamma_{AB}%
^{R}<q\alpha$ there will be a region in the phase space $\alpha,\beta$
(shrinking as $\Gamma_{AB}^{L},\Gamma_{AB}^{R}$ increase) where the symmetry
broken phase will exist.

\acknowledgments We thank D. Mukamel, R. Willmann and C. Bahadoran for fruitful 
discussions. V.P. thanks Deutsche Forschungsgemeinschaft for financial support.

%\newpage

%\bibliographystyle{revsymb}
\bibliographystyle{apsrev}
\bibliography{ABO}

\begin{thebibliography}{25}
\expandafter\ifx\csname natexlab\endcsname\relax\def\natexlab#1{#1}\fi
\expandafter\ifx\csname bibnamefont\endcsname\relax
  \def\bibnamefont#1{#1}\fi
\expandafter\ifx\csname bibfnamefont\endcsname\relax
  \def\bibfnamefont#1{#1}\fi
\expandafter\ifx\csname citenamefont\endcsname\relax
  \def\citenamefont#1{#1}\fi
\expandafter\ifx\csname url\endcsname\relax
  \def\url#1{\texttt{#1}}\fi
\expandafter\ifx\csname urlprefix\endcsname\relax\def\urlprefix{URL }\fi
\providecommand{\bibinfo}[2]{#2}
\providecommand{\eprint}[2][]{\url{#2}}

\bibitem[{\citenamefont{Schütz}(2000)}]{Schu00}
\bibinfo{author}{\bibfnamefont{G.~M.} \bibnamefont{Schütz}},
  \emph{\bibinfo{title}{Exactly solvable models for many-body systems far from
  equilibrium}} (\bibinfo{publisher}{Academic Press, London},
  \bibinfo{year}{2000}), \bibinfo{note}{in: Phase Transitions and Critical
  Phenomena, ed. C.Domb and J.L. Lebowitz, Vol. 19}.

\bibitem[{\citenamefont{Liggett}(1999)}]{Liggett1999}
\bibinfo{author}{\bibfnamefont{T.~M.} \bibnamefont{Liggett}},
  \emph{\bibinfo{title}{Stochastic interacting systems: contact, voter and
  exclusion processes}} (\bibinfo{publisher}{Springer, Berlin},
  \bibinfo{year}{1999}).

\bibitem[{\citenamefont{Schmittmann and Zia}(1995)}]{Zia}
\bibinfo{author}{\bibfnamefont{B.}~\bibnamefont{Schmittmann}} \bibnamefont{and}
  \bibinfo{author}{\bibfnamefont{R.~K.~P.} \bibnamefont{Zia}},
  \emph{\bibinfo{title}{Statistical Mechanics of Driven Diffusive Systems}}
  (\bibinfo{publisher}{Academic Press, London}, \bibinfo{year}{1995}),
  \bibinfo{note}{in: Phase Transitions and Critical Phenomena, ed. C.Domb and
  J.L. Lebowitz, Vol. 17}.

\bibitem[{\citenamefont{Krug}(1991)}]{Krug91}
\bibinfo{author}{\bibfnamefont{J.}~\bibnamefont{Krug}}, \bibinfo{journal}{Phys.
  Rev. Lett.} \textbf{\bibinfo{volume}{67}}, \bibinfo{pages}{1882}
  (\bibinfo{year}{1991}).

\bibitem[{\citenamefont{Schütz and Domany}(1993)}]{ASEP1}
\bibinfo{author}{\bibfnamefont{G.}~\bibnamefont{Schütz}} \bibnamefont{and}
  \bibinfo{author}{\bibfnamefont{E.}~\bibnamefont{Domany}}, \bibinfo{journal}{J
  Stat. Phys} \textbf{\bibinfo{volume}{72}}, \bibinfo{pages}{277}
  (\bibinfo{year}{1993}).

\bibitem[{\citenamefont{Derrida et~al.}(1993)\citenamefont{Derrida, Evans,
  Hakim, and Pasquier}}]{ASEP}
\bibinfo{author}{\bibfnamefont{B.}~\bibnamefont{Derrida}},
  \bibinfo{author}{\bibfnamefont{M.~R.} \bibnamefont{Evans}},
  \bibinfo{author}{\bibfnamefont{V.}~\bibnamefont{Hakim}}, \bibnamefont{and}
  \bibinfo{author}{\bibfnamefont{V.}~\bibnamefont{Pasquier}},
  \bibinfo{journal}{J Phys. A} \textbf{\bibinfo{volume}{26}},
  \bibinfo{pages}{1493} (\bibinfo{year}{1993}).

\bibitem[{\citenamefont{Kolomeisky et~al.}(1998)\citenamefont{Kolomeisky,
  Schütz, Kolomeisky, and Straley}}]{Kolo98}
\bibinfo{author}{\bibfnamefont{A.~B.} \bibnamefont{Kolomeisky}},
  \bibinfo{author}{\bibfnamefont{G.~M.} \bibnamefont{Schütz}},
  \bibinfo{author}{\bibfnamefont{E.~B.} \bibnamefont{Kolomeisky}},
  \bibnamefont{and} \bibinfo{author}{\bibfnamefont{J.~P.}
  \bibnamefont{Straley}}, \bibinfo{journal}{J. Phys. A}
  \textbf{\bibinfo{volume}{31}}, \bibinfo{pages}{6911} (\bibinfo{year}{1998}).

\bibitem[{\citenamefont{Popkov and Schütz}(1999)}]{Gunter_Slava_Europhys}
\bibinfo{author}{\bibfnamefont{V.}~\bibnamefont{Popkov}} \bibnamefont{and}
  \bibinfo{author}{\bibfnamefont{G.~M.} \bibnamefont{Schütz}},
  \bibinfo{journal}{Europhys. Lett} \textbf{\bibinfo{volume}{48}},
  \bibinfo{pages}{257} (\bibinfo{year}{1999}).

\bibitem[{\citenamefont{Evans et~al.}(1995)\citenamefont{Evans, Foster,
  Godr\`eche, and Mukamel}}]{Mukamel95}
\bibinfo{author}{\bibfnamefont{M.~R.} \bibnamefont{Evans}},
  \bibinfo{author}{\bibfnamefont{D.~P.} \bibnamefont{Foster}},
  \bibinfo{author}{\bibfnamefont{C.}~\bibnamefont{Godr\`eche}},
  \bibnamefont{and} \bibinfo{author}{\bibfnamefont{D.}~\bibnamefont{Mukamel}},
  \bibinfo{journal}{J. Stat. Phys.} \textbf{\bibinfo{volume}{80}},
  \bibinfo{pages}{898} (\bibinfo{year}{1995}).

\bibitem[{\citenamefont{Popkov and Peschel}(2001)}]{Peschel}
\bibinfo{author}{\bibfnamefont{V.}~\bibnamefont{Popkov}} \bibnamefont{and}
  \bibinfo{author}{\bibfnamefont{I.}~\bibnamefont{Peschel}},
  \bibinfo{journal}{Phys. Rev. E} \textbf{\bibinfo{volume}{64}},
  \bibinfo{pages}{026126} (\bibinfo{year}{2001}).

\bibitem[{\citenamefont{Levine and Willmann}(2004)}]{Erel}
\bibinfo{author}{\bibfnamefont{E.}~\bibnamefont{Levine}} \bibnamefont{and}
  \bibinfo{author}{\bibfnamefont{R.}~\bibnamefont{Willmann}},
  \bibinfo{journal}{J. Phys. A} \textbf{\bibinfo{volume}{37}},
  \bibinfo{pages}{3333} (\bibinfo{year}{2004}).

\bibitem[{\citenamefont{Rakos and Sch{\"u}tz}(2004)}]{Rako04a}
\bibinfo{author}{\bibfnamefont{A.}~\bibnamefont{Rakos}} \bibnamefont{and}
  \bibinfo{author}{\bibfnamefont{G.~M.} \bibnamefont{Sch{\"u}tz}},
  \bibinfo{journal}{J. Stat. Phys.} p. \bibinfo{pages}{to appear}
  (\bibinfo{year}{2004}).

\bibitem[{\citenamefont{Antal and Schütz}(2000)}]{TiborAntal}
\bibinfo{author}{\bibfnamefont{T.}~\bibnamefont{Antal}} \bibnamefont{and}
  \bibinfo{author}{\bibfnamefont{G.~M.} \bibnamefont{Schütz}},
  \bibinfo{journal}{Phys. Rev. E.} \textbf{\bibinfo{volume}{62}},
  \bibinfo{pages}{83} (\bibinfo{year}{2000}).

\bibitem[{Bah()}]{Baha04}
\bibinfo{note}{C. Bahadoran: Hydrodynamics of asymmetric particle systems with
  open boundaries, in preparation}.

\bibitem[{\citenamefont{Schütz}(2003)}]{Gunter_reviewJPA}
\bibinfo{author}{\bibfnamefont{G.~M.} \bibnamefont{Schütz}},
  \bibinfo{journal}{J. Phys. A} \textbf{\bibinfo{volume}{36}},
  \bibinfo{pages}{R339} (\bibinfo{year}{2003}).

\bibitem[{\citenamefont{Popkov}(2004)}]{reflections_JPA}
\bibinfo{author}{\bibfnamefont{V.}~\bibnamefont{Popkov}}, \bibinfo{journal}{J.
  Phys. A} \textbf{\bibinfo{volume}{37}}, \bibinfo{pages}{1545}
  (\bibinfo{year}{2004}).

\bibitem[{\citenamefont{Arndt et~al.}(1998)\citenamefont{Arndt, Heinzel, and
  Rittenberg}}]{Rittenberg98}
\bibinfo{author}{\bibfnamefont{P.~F.} \bibnamefont{Arndt}},
  \bibinfo{author}{\bibfnamefont{T.}~\bibnamefont{Heinzel}}, \bibnamefont{and}
  \bibinfo{author}{\bibfnamefont{V.}~\bibnamefont{Rittenberg}},
  \bibinfo{journal}{J. Phys. A} \textbf{\bibinfo{volume}{31}},
  \bibinfo{pages}{831} (\bibinfo{year}{1998}).

\bibitem[{\citenamefont{Popkov and Schütz}(2003)}]{GunterSlava_StatPhys}
\bibinfo{author}{\bibfnamefont{V.}~\bibnamefont{Popkov}} \bibnamefont{and}
  \bibinfo{author}{\bibfnamefont{G.~M.} \bibnamefont{Schütz}},
  \bibinfo{journal}{J Stat. Phys.} \textbf{\bibinfo{volume}{112}},
  \bibinfo{pages}{523} (\bibinfo{year}{2003}).

\bibitem[{\citenamefont{Fritz and T{\'o}th}(2004)}]{FritzToth}
\bibinfo{author}{\bibfnamefont{J.}~\bibnamefont{Fritz}} \bibnamefont{and}
  \bibinfo{author}{\bibfnamefont{B.}~\bibnamefont{T{\'o}th}},
  \bibinfo{journal}{Comm. Math. Phys.} \textbf{\bibinfo{volume}{249}},
  \bibinfo{pages}{1} (\bibinfo{year}{2004}).

\bibitem[{\citenamefont{Serre}(2000)}]{Serre2000}
\bibinfo{author}{\bibfnamefont{D.}~\bibnamefont{Serre}},
  \emph{\bibinfo{title}{Systems of conservation laws, Vol.2}}
  (\bibinfo{publisher}{Cambridge University Press}, \bibinfo{year}{2000}).

\bibitem[{\citenamefont{Popkov and Salerno}(2004)}]{Mario_multiASEP}
\bibinfo{author}{\bibfnamefont{V.}~\bibnamefont{Popkov}} \bibnamefont{and}
  \bibinfo{author}{\bibfnamefont{M.}~\bibnamefont{Salerno}},
  \bibinfo{journal}{Phys. Rev. E} \textbf{\bibinfo{volume}{69}},
  \bibinfo{pages}{046103} (\bibinfo{year}{2004}).

\bibitem[{\citenamefont{Godr\`eche et~al.}(1995)\citenamefont{Godr\`eche, Luck,
  Evans, Mukamel, Sandow, and Speer}}]{MukamelToyModel}
\bibinfo{author}{\bibfnamefont{C.}~\bibnamefont{Godr\`eche}},
  \bibinfo{author}{\bibfnamefont{J.~M.} \bibnamefont{Luck}},
  \bibinfo{author}{\bibfnamefont{M.~R.} \bibnamefont{Evans}},
  \bibinfo{author}{\bibfnamefont{D.}~\bibnamefont{Mukamel}},
  \bibinfo{author}{\bibfnamefont{S.}~\bibnamefont{Sandow}}, \bibnamefont{and}
  \bibinfo{author}{\bibfnamefont{E.~R.} \bibnamefont{Speer}},
  \bibinfo{journal}{J. Phys. A} \textbf{\bibinfo{volume}{28}},
  \bibinfo{pages}{6} (\bibinfo{year}{1995}).

\bibitem[{\citenamefont{Lahiri et~al.}(2000)\citenamefont{Lahiri, Barma, and
  Ramaswamy}}]{Lahi00}
\bibinfo{author}{\bibfnamefont{R.}~\bibnamefont{Lahiri}},
  \bibinfo{author}{\bibfnamefont{M.}~\bibnamefont{Barma}}, \bibnamefont{and}
  \bibinfo{author}{\bibfnamefont{S.}~\bibnamefont{Ramaswamy}},
  \bibinfo{journal}{Phys. Rev. E} \textbf{\bibinfo{volume}{61}},
  \bibinfo{pages}{1648} (\bibinfo{year}{2000}).

\bibitem[{\citenamefont{Das et~al.}(2001)\citenamefont{Das, Basu, Barma, and
  Ramaswamy}}]{Das01}
\bibinfo{author}{\bibfnamefont{D.}~\bibnamefont{Das}},
  \bibinfo{author}{\bibfnamefont{A.}~\bibnamefont{Basu}},
  \bibinfo{author}{\bibfnamefont{M.}~\bibnamefont{Barma}}, \bibnamefont{and}
  \bibinfo{author}{\bibfnamefont{S.}~\bibnamefont{Ramaswamy}},
  \bibinfo{journal}{Phys. Rev. E} \textbf{\bibinfo{volume}{64}},
  \bibinfo{pages}{021402} (\bibinfo{year}{2001}).

\bibitem[{\citenamefont{T{\'o}th and Valk{\'o}}(2003)}]{Valk03}
\bibinfo{author}{\bibfnamefont{B.}~\bibnamefont{T{\'o}th}} \bibnamefont{and}
  \bibinfo{author}{\bibfnamefont{B.}~\bibnamefont{Valk{\'o}}},
  \bibinfo{journal}{J Stat. Phys.} \textbf{\bibinfo{volume}{112}},
  \bibinfo{pages}{497} (\bibinfo{year}{2003}).

\end{thebibliography}
%Produces the bibliography via BibTeX.
%
\newpage

%TCIMACRO{\FRAME{ftbpFU}{5.0548in}{3.5405in}{0pt}{\Qcb{Domain wall reflection
%from the left boundary, as given by Monte Carlo simulations. Particles were
%initially distributed randomly without correlations with constant densities
%$\rho^{A}=0.7,\rho^{B}=0.1$, fitting the right boundary. The average profiles
%after $300$ Monte Carlo steps are shown, and compared with the hydrodynamic
%limit (\ref{hydro}) evolution (continuous curves). Averaging over $5\ast
%10^{5}$ histories is made. The smoothing of the Monte-Carlo density profile is
%due to fluctuations in the shock position which is scaled out in hydrodynamic
%description. }}{\Qlb{fig_MC_Lrefl}}{fig_abo_mc_lrefl.eps}%
%width 5.0548in;  height 3.5405in;  depth 0pt;  original-width 5.0004in;
%original-height 3.4938in;  cropleft "0";  croptop "1";  cropright "1";
%filename 'ABO_fig/fig_ABO_MC_Lrefl.eps';file-properties "XNPEU";}}}%
%BeginExpansion

%EndExpansion
%

%TCIMACRO{\FRAME{ftbpFU}{5.0548in}{3.5405in}{0pt}{\Qcb{Domain wall reflection
%from the right boundary. Particles were initially distributed randomly without
%correlations with constant densities $\rho^{A}=0.25,\rho^{B}=0.55$ fitting the
%left boundary density. Average density profiles after $t=300$ MCS steps are
%shown. The boundary densities are as on Fig. \ref{fig_MC_Lrefl}. Lines show
%the hydrodynamic evolution (\ref{hydro}) for the two particle species. The
%smoothing of the Monte-Carlo density profile is due to fluctuations in the
%shock position which is scaled out in hydrodynamic description.}%
%}{\Qlb{fig_MC_Rrefl}}{fig_abo_mc_rrefl.eps}%
%width 5.0548in;  height 3.5405in;  depth 0pt;  original-width 5.0004in;
%original-height 3.4938in;  cropleft "0";  croptop "1";  cropright "1";
%filename 'ABO_fig/fig_ABO_MC_Rrefl.eps';file-properties "XNPEU";}}}%
%BeginExpansion

%EndExpansion

%

%TCIMACRO{\FRAME{ftbpFU}{5.0548in}{3.5405in}{0pt}{\Qcb{Left reflection map for
%the left boundary densities $0.25,0.55$ of $A$ and $B$ particles respectively.
%The crossmark labelled $L$ marks the point of the left boundary density. The
%curve containing this point $L$ is a location of all possible densities
%$r^{A},r^{B}$ of the \ reflected waves, except at the region HD. The straight
%lines show initial bulk densities with the same outcome of a reflection (point
%of intersection of the straight line with the curve). In HD region the initial
%wave stays glued to the left boundary. In the RW region (the upper triangle)
%the reflection results in a rarefaction wave (scaling as $(x/t)$ with time),
%converging to point $L$. The dotted line $r^{A}+r^{B}=1$ shows boundaries of
%physical region.}}{\Qlb{fig_L}}{fig_l.eps}%
%width 5.0548in;  height 3.5405in;  depth 0pt;  original-width 5.0004in;
%original-height 3.4938in;  cropleft "0";  croptop "1";  cropright "1";
%cropbottom "0";  filename 'ABO_fig/fig_L.eps';file-properties "XNPEU";}}}%
%BeginExpansion

%EndExpansion%
%TCIMACRO{\FRAME{ftbpFU}{5.0548in}{3.5405in}{0pt}{\Qcb{Right reflection map for
%the right boundary density $0.1,0.7$ of $A$,$B$ particles. The crossmark
%labelled $R$ marks the point of the right boundary density. The thick curve is
%a location of all possible densities $r^{A},r^{B}$ of the \ reflected waves,
%except at the upper left triangle and in region RB. Straight lines show
%initial bulk densities with the same outcome of a reflection (point of
%intersection of the line with the thick curve). In the upper corner the
%initial wave stays glued to the right boundary. In the region RB an
%intermediate plateau appears, with the densities corresponding to the broken
%line passing through point $R$. The final result of this reflection is a wave
%matching the right boundary $R$. The dotted line $r^{A}+r^{B}=1$ shows
%boundaries of physical region.}}{\Qlb{fig_R}}{fig_r.eps}%
%width 5.0548in;  height 3.5405in;  depth 0pt;  original-width 5.0004in;
%original-height 3.4938in;  cropleft "0";  croptop "1";  cropright "1";
%cropbottom "0";  filename 'ABO_fig/fig_R.eps';file-properties "XNPEU";}}}%
%BeginExpansion
%EndExpansion
%

%TCIMACRO{\FRAME{ftbpFU}{5.0548in}{3.5405in}{0pt}{\Qcb{ Curves of the right and
%left boundary reflection combined together. The intersection $S$, is the
%stationary state density achieved through the infinite number of reflections.
%The crossmarks labelled by $R$ and $L$ indicate the left and right boundary
%densities, and the cross at the intersection the stationary density obtained
%by Monte Carlo simulation of the system of $300$ sites. The system was
%equilibrated for = $4\ast10^{5}$ Monte Carlo Steps (MCS), after which the
%averaging over again $4\ast10^{5}$ MCS and $10$ different histories was done.
%The big star to the righthand side of $S$ marks the result of single
%reflection for an initial profile fitting the left boundary density point $L$
%(from Monte Carlo calculations).}}{\Qlb{fig_RL}}{fig_abo_map.eps}%
%width 5.0548in;  height 3.5405in;  depth 0pt;  original-width 5.0004in;
%original-height 3.4938in;  cropleft "0";  croptop "1";  cropright "1";
%cropbottom "0";  filename 'ABO_fig/fig_ABO_Map.eps';file-properties "XNPEU";}%
%}}%
%BeginExpansion
%EndExpansion%
%TCIMACRO{\FRAME{ftbpFU}{4.1303in}{4.0577in}{0pt}{\Qcb{ Approach to the
%stationary point $S$ through an infinite sequence of reflections. The filled
%circles show the location of subsequent domain wall densities $r^{A},r^{B}$.
%The circles $k,k+2,k+4$ correspond to the result of the left reflection, while
%the other circles correspond to the right reflection.}}{\Qlb{fig_AB0_enlarge}%
%}{fig_ab0_enlarge.eps}{\special{ language "Scientific Word";  type "GRAPHIC";
%width 4.1303in;  height 4.0577in;  depth 0pt;  original-width 4.0802in;
%original-height 4.0075in;  cropleft "0";  croptop "1";  cropright "1";
%filename 'ABO_fig/fig_AB0_enlarge.eps';file-properties "XNPEU";}}}%
%BeginExpansion
%EndExpansion%
%TCIMACRO{\FRAME{ftbpFU}{5.0548in}{3.5405in}{0pt}{\Qcb{ Curves of the right
%(filled triangles) and left (empty triangles) boundary reflection for the
%symmetric setting (\ref{symmetric_boundary_conditions}), as given by numerical
%integration of the hydrodynamic limit equations (\ref{hydro}). Triangles
%labelled $R$ and $L$ indicate left and right boundary densities, and the point
%$S$ the stationary density, achieved through infinite series of reflections.
%The cross at the the intersection point $S$ marks the stationary density
%obtained by Monte Carlo simulation of the system of $300$ sites. The broken
%and dotted lines intersecting at $S$ show characteristic curves at point $S$
%(see (\ref{sol_L}),(\ref{sol_R})). }}{\Qlb{fig_symm_Map}}{fig_symm_map.eps}%
%width 5.0548in;  height 3.5405in;  depth 0pt;  original-width 5.0004in;
%original-height 3.4938in;  cropleft "0";  croptop "1";  cropright "1";
%cropbottom "0";  filename 'ABO_fig/fig_symm_Map.eps';file-properties "XNPEU";}%
%}}%
%BeginExpansion
%EndExpansion

%title = "Derivation of the Leroux system as the hydrodynamic limit 
%of a two-component lattice gas, math.PR/0304481 ",

\end{document}